# Orchestrating the Twin Transition in Multinational Corporations: Technology Roadmapping for Green and Digital Global Business Services

Han-Teng Liao
*Independent Researcher*

Penang, Malaysia
h.liao@ieee.org
0000-0003-1081-5599

Karen Ang
*Independent Researcher*

Kulim, Malaysia
karenang118@gmail.com
0009-0008-5923-0106

***Abstract*—Global Business Services (GBS) have emerged as a "living laboratory" for the Twin Transition of Green and Digital Transformation, as multinational corporations (MNCs) face increasing pressure to harmonize digital efficiency with environmental stewardship. Aiming to derive a socio-technical framework, this paper synthesizes Technology Roadmapping (TRM) with the International Telecommunication Union (ITU) ICT-centric innovation ecosystem toolkit. A bibliometric analysis of research clusters reveals an evolutionary shift from basic process automation toward "Sustainable Intelligence," identifying the GBS unit as a central "operational airlock" that mediates between landscape pressures—such as the EU's dual mandate and Carbon Border Adjustment Mechanisms—and niche innovations in AI-native workflows. The study further maps these clusters onto a stakeholder engagement canvas, highlighting how resilient "Middle Power" hubs in Poland, Portugal, and Malaysia are bypassing the middle-income trap to provide a "third way" for global value chains amidst a bifurcated geopolitical cloud. The results offer a data-driven design approach for leaders and entrepreneurial support networks to orchestrate talent and supply chain flows, thereby enriching the conceptual understanding of Industry 5.0 and the role of GBS as a primary mechanism for navigating a volatile, multipolar digital economy.**

## I. INTRODUCTION

Global service delivery is undergoing a structural realignment toward "digital resonance"—the capacity for high-tech automation and cybersecurity—, superseding traditional labor arbitrage as the new driver of location competitiveness [1]. Artificial intelligence (AI) chatbots likely have impacts on professional services [2]. Consequently, Global Business Services (GBS) are vital "living laboratories" for the corporate Twin Transition, harmonizing technical execution with sustainability goals. Nations such as Portugal (Green Tech Center of Excellence, or CoE), Poland (EMEA AI-native hub), and Malaysia (GBS5.0 Asia-Pacific digital gateway) illustrate real-world ecosystem orchestration beyond the well-known cases of India and Philippines. These hubs now operationalize the Twin Transition at scale. Table I details these illustrative ecosystems as of early 2026, showing how they —collectively employing nearly one million professionals— have upgraded into national innovation ecosystems using Information and Communication Technologies (ICTs): Poland as the "Silicon Valley of Central Europe," Portugal as one AI-native hub, Malaysia Penang as the "Silicon Valley of the East".

TABLE I. 2026 GBS LANDSCAPE: ECOSYSTEM MATURITY

| Country | Reported or Estimated Data | | Strategic Focus (2026) | Primary Agency |
|---|---|---|---|---|
| | *Centers* | *Workforce* | | |
| Poland | ~2,000 | 488,700+ | AI-Native Workflows, Fintech | ABS [a] |
| Portugal | ~270 | 93,000+ | Green Tech; Software CoE | AICEP [b] |
| Malaysia | ~750+ | 250,000+ | GBS 5.0; high-value digital GBS hub | MDEC [c] |

a. Association of Business Service Leaders (ABSL) [3]: The leading organization representing the Modern Business Services sector in Poland, providing the primary annual census and benchmarking data used by the Polish government and global investors.

b. Agência para o Investimento e Comércio Externo de Portugal (AICEP) [1]: A government entity focused on encouraging foreign direct investment (FDI) and promoting Portuguese exports. AICEP manages the GBS "Site Selection" platform and coordinates the "Centers of Excellence" (CoE) strategy to attract sustainable, tech-heavy hubs.

c. Malaysia Digital Economy Corporation (MDEC) [4]: The architects of the "Malaysia Digital" (MD) status and the GBS 5.0 framework; a government agency under the Ministry of Digital (Malaysia) tasked with leading the digital economy.

The year 2026 marks a critical inflection point: the obsolescence of "lift and shift" models rooted in labor arbitrage. Industry observations suggest a pivot toward "lean orchestration," where process normalization is a mandatory prerequisite for AI integration [5]. However, this transition is governed by the "human factor": As a capability multiplier, its efficacy is often throttled by a "leadership gap," wherein executive ROI expectations clash with organizational inertia.

As a result, the GBS value proposition is shifting from transactional execution to strategic insourcing, where cultural resilience and high-value talent remain competitive advantages that technology alone cannot replicate [5]. This shift intersects with "middle power" geopolitics. Emerging hubs—Poland's 488,700-strong AI-native workforce, Portugal's Green Tech CoE, and Malaysia's GBS 5.0 Digital Gateway—function as "operational airlocks," insulating global value chains (GVC) from U.S.-China bifurcation while mitigating middle-income traps through high-value talent and resilient supply chain management.

Successful orchestration requires harmonizing technical dimensions like AI and Green ICT with escalating stakeholder engagement: from informing and consulting, to collaborating and co-creating [6]. Heavy infrastructure investment in hubs like Penang, has enabled high-value functions such as IC design and R&D [7], and yet U.S.-China semiconductor rivalry raises questions about long-term sustainability. While 84% of Portuguese centers are already upskilling for AI and ESG requirements [8], it remains unclear whether "intelligence-driven" upgrades can broadly mitigate automation risks or will merely concentrate value among elite technical talents. A systematic review or a roadmapping exercise is thus required to determine if and how academic discourse reflects this shift toward high-value orchestration — or remains tethered to cost-centric paradigms.

## II. Existing Theories & Previous Work

The structural realignment of global value chains (GVC) requires a theoretical perspective that explains how former transactional outsourcing hubs mature into highly integrated innovation ecosystems. This review evaluates the cross-disciplinary convergence of GBS evolution, STS theory, and the MLP-STT to baseline digital and green orchestration.

### *A. From Data Integration to Digital Orchestration*

Tracing the GBS trajectory reveals a shift from localized cost reduction toward distributed, sustainable ecosystems. Early centralized delivery relied on integrated information systems, such as NEC's Ateam platform, to unify global administrative and development functions [9]. From an evolutionary STT perspective, these early practices illustrate the transition from concurrent engineering toward cross-border competitive advantage [10] and the orchestration of enterprise ecosystems [11], and "Sustainable Intelligence."

While classical outsourcing prioritized cost, virtual production networks now induce systemic barriers and opportunities for emerging economies. Global digital platforms impose notable constraints on worker livelihoods and localized economic upgrading [12]. They often function as offshoring institutions that reshape transaction costs while institutionalizing power asymmetries between multinational corporations (MNCs) and local providers [13]. This drives a distinct global polarization that concentrates high-value tasks within mature hubs while trapping peripheral nodes [14].

To counter this, contemporary GBS centers rely on "dual embeddedness," navigating both internal corporate networks and local host-country digital ecosystems [15]. Amid shifting geopolitics, countries like Poland, Portugal, and Malaysia utilize these GBS policies to mitigate "middle-income traps" through high-value talent and resilient supply chain management. Analyzing this transformation is essential for optimizing ecosystem innovation strategies that align complex technical, social, and geopolitical interests [16]. However, such alignment is frequently obstructed by "legacy drag"—the cumulative weight of non-standardized processes inherited from the labor-arbitrage era.

### *B. Automation, Optimization, and the "Maturity Trap"*

As firms transitioned toward more sophisticated service models from labor arbitrage to value-driven digital orchestration, Robotic Process Automation (RPA) emerged as a primary strategic lever [17]. These developments were expected to drive the digital transformation of the service sector, facilitating more agile business environments [18], [19]. Empirical research across diverse geographies—including Poland [20], [21], [22] and China [23], [24] — validates these shifts. To ensure the outcomes of such efforts, maturity frameworks like the KPMG five-stage GBS model were developed to guide organizations from fragmented operations to an integrated service model [25]. Critical dimensions identified in these models—**service portfolio, talent management, technology enablement, risk management, and performance management**—suggest a primary academic interest in technology management for narrower organizational outcomes.

In practice, many GBS organizations remain caught in a "maturity trap," where AI-native ambitions are retrofitted onto fragmented, legacy architectures. As Lubaczewski [5] notes, without radical process normalization, these frameworks act as a "multiplier of inefficiency" rather than a catalyst for value.

### *C. Socio-Technical Systems and the Digitalization Paradox*

Instead of assuming a "greenfield" environment, viewing modern GBS hubs as complex Socio-Technical Systems (STS)—where the "Technical" (AI-native workflows) harmonizes with the "Social" (human-centric Industry 5.0, talent development, and trade policies)—captures maturity trap realities under green and digital transformation demands. While GBS evolution historically emphasized technical efficiency, recent studies in digital public services suggest a "digitalization paradox," where advanced digital tools can inversely affect stakeholder well-being [26]. To mitigate this, smart digital platforms must prioritize open innovation and carbon-neutral management [27]. Consequently, management must treat "messy legacy processes" not as a technical glitch, but as a systemic socio-technical transitions to adapt to technical velocity and social demands, requiring a System of Systems (SoS) approach [28]. This underscores the necessity of an MLP-STT approach in GBS, which prioritizes process hygiene, environmental regimes, and leadership alignment, especially in AI-driven workflows involving ESG targets.

### *D. Multi-Level Perspective (MLP) and the Twin Transition*

The Multi-Level Perspective (MLP) framework analyzes socio-technical transitions across three nested levels of increasing structuration: micro-level niches (the locus of radical innovations), meso-level socio-technical regimes (stable systems of rules and practices), and macro-level landscapes (exogenous global pressures on regimes)[29], [30], [31], [32], [33].

Despite the robustness of MLP and STT research, its application to the GBS domain remains largely unexplored. Current GBS literature focuses heavily on internal maturity models but often overlooks exogenous "Landscape" pressures—such as the European Union's "Twin Transition" mandates—that destabilize existing corporate "Regimes."

While traditional Global Business Services (GBS) and Global Mobility Services (GMS) models have prioritized labor and immigration compliance [11], this transition now demands the integration of agentic workflows with AI. Historically, MNCs have maintained technical and information asymmetries that hinder holistic change; thus, a conceptual framework utilizing STT and Industry 5.0 is required to democratize innovation flows.

AI-enabled services for Sustainable Development Goals (SDGs) provide the necessary "Niche" evidence for GBS to transition from a cost-center to a sustainability-driver. Addressing this gap requires moving beyond static maturity assessments toward an approach that captures the active interplay between niche innovations and regime stability. Accordingly, this study adopts a dual-horizon research design—detailed in the following methodology—to treat GBS as a living laboratory for the socio-technical transition of AI and human resources.

## III. Methodology: Technology Roadmapping and Bibliometrics to Identify Niches and Regimes

The research design integrates *retrospective science mapping* — including longitudinal bibliometric analysis — with *prospective service design* to architect the future green and digital GBS. This dual-horizon approach bridges established socio-technical research clusters with forward-looking ecosystem innovation technology roadmapping, situating GBS as the transformative 'operational airlock' within GVC.

### A. Research Questions and Hypothesis

The central research question driving this study is: *How has the GBS domain evolved from a cost-arbitrage model toward an ecosystem innovation governance mechanism, and what socio-technical dynamics can be leveraged to drive the transition of safer and sustainable GVC?*

Informed by the MLP and STT research, two subordinate questions sharpen the inquiry. First, which thematic clusters in the GBS literature constitute stabilized "regimes" — well-developed, centrally connected bodies of practice — and which remain "niche" innovations with high internal density but limited cross-cluster diffusion? Second, how can Technology Roadmapping (TRM), informed by bibliometric evidence, provide GBS leaders with an actionable framework to anticipate the absorption of current niches — such as ESG, Blockchain, and Employee Health — into future regimes?

The study hypothesizes that GBS has evolved into a sophisticated "living laboratory" for the socio-technical transition of AI and human resources. Thematic evolution analysis, when mapped onto MLP and STT theory, can operationalize the niche-regime distinction as a predictive instrument for technology roadmapping. Specifically, *Niche* themes surfaced from bibliometric data correspond to MLP niche innovations accumulating internal coherence and approaching the threshold at which they exert landscape-level pressure on the dominant regime. Conversely, *Motor* and *Basic* themes reveal the stabilized socio-technical regime — the rules, practices, and expectations structuring incumbent GBS operations. Bibliometric data thus provides an empirically grounded basis for roadmapping, distinguishing signals worth investing in from noise.

### B. Applied Research Methods, Models, and Instruments

This study employs a mixed-methods two-step approach. The first step applies longitudinal bibliometric analysis to map the conceptual and social structure of the GBS knowledge domain across 1993–2026. The second step integrates these findings with the International Telecommunication Union (ITU) ICT-Centric Innovation Ecosystem Toolkit to translate empirical clusters into actionable stakeholder canvases.

Technology Roadmapping (TRM) practice has responded to disruptive technologies and social changes for over three decades, proving flexible and well-documented utility [34], [35]. Integrating bibliometric data enables TRM to combine seamlessly with qualitative methodologies and visual artifacts to generate competitive intelligence [36], [37], [38], while design thinking processes aid problem-solving within complex ecosystem processes [39]. In the context of the Twin Transition, recent scientometric work has successfully developed roadmapping taxonomies [40], [41], partner maps [42], and a smart carbon platform design based on ITU and Chinese standards [27].

By unifying *retrospective science mapping* with *prospective service design* through the MLP lens of FBS, the study observes how transactional GBS practices respond to landscape pressures—such as carbon neutrality and AI disruption—by fostering niche innovations. This highlights how governments and entrepreneurial support networks, such as ABSL (Poland), MDEC (Malaysia), and AICEP (Portugal), act as regime-influencing forces that compel Multinational Corporations (MNCs) to adapt their strategic roadmaps.

### C. Data collection

For *retrospective science mapping,* a targeted query was executed in the Web of Science database in February 2026, capturing 979 documents spanning 1993 to 2026:

TS = ("Global Business Service*" OR "Shared Service Cent*" OR "business process outsourcing" OR "business process optimi*" ) OR TS = ( ( "Service Industries" OR "Service Economy" OR "Service Design" OR "Service Computing" ) AND ( "digital economy" OR "platform economy" OR "world economy" OR "Multinational Corporation*") ) and Article or Proceeding Paper (Document Types)

To ensure taxonomic accuracy [43], [39], a specialized thesaurus was compiled to merge synonyms (e.g., "shared service centres" and "SSC") and differentiate strategically distinct terms (e.g., business process optimization vs. business process outsourcing) that share the BPO abbreviation and risk conflation without explicit disambiguation.

### D. Analysis Methods

For *retrospective science mapping,* a bibliometric analysis was conducted using a combination of Bibliometrix (R-package) [44], VOSviewer, and customized Python script, with the aim to reveal the conceptual and social structures of the GBS knowledge domain. Science mapping of relevant concepts and contributing countries provides critical intelligence on "research fronts", offering insights into how information and communication flows impact the orchestration between digital execution and sustainability goals. The visualization informs a structured taxonomy, allowing leaders to align the technical dimensions of GBS with the Environmental, Social, and Governance (ESG) requirements of Industry 5.0, thereby overcoming the "maturity trap" and achieving ecosystem innovation [16].

Critically, the thematic evolution analysis — operationalized through three strategic maps spanning 1993–2026 — serves a dual function. It both describes the historical trajectory of the GBS field and generates forward-looking niche detection intelligence for technology roadmapping. Themes migrating from the *Niche* quadrant toward *Motor* or *Basic* status across successive periods represent socio-technical transitions in progress: the MLP equivalent of niche innovations accumulating sufficient momentum to destabilize the incumbent regime. This will reveal trajectory of terms as directly actionable for GBS leaders constructing five-year roadmaps.

### E. Innovation Ecosystem Service Design

For *prospective service design,* the *retrospective science mapping* findings are then synthesized using the design vocabulary and canvases detailed in the ITU ICT-Centric Innovation Ecosystem Toolkit [6]. As a globally recognized framework to design sustainable, multi-stakeholder digital projects, its diagnostic canvases map project requirements against regional constraints and stakeholder configurations.

Table II contextualizes these tools within the MLP framework, which analyzes socio-technical transitions. Applied to GBS as a living laboratory for the socio-technical transition of AI and human resources, each tool addresses a distinct MLP level. The *Stakeholder Engagement Tool* maps the current regime by capturing common stakeholder needs, constraints, and value propositions. The *Ecosystem Canvas Tool* aligns niche and regime actors around a shared manifesto of roles and responsibilities. The *Ecosystem Maturity Map* tracks progression from niche experimentation to regime consolidation at national or city level. The *Sector Development Canvas* monitors cross-sectoral supply chain and talent flows as green and digital transformation reshape sectoral boundaries. The *Storytelling Tool* sustains stakeholder relationships across transition phases, maintaining narrative coherence as regimes evolve.

TABLE II. EXPANDED DESIGN TOOLS BASED ON ITU'S ICT-CENTRIC ECOSYSTEM TOOLKIT FOR MLP-STT SYSTEM THINKING

| Tool | Goals |
|---|---|
| Stakeholder Engagement Tool | • Capture common needs and the "regime" environment<br>• Establish value propositions among stakeholders' relationships |
| Ecosystem Canvas Tool | • Align interests and actions based on the understanding of the socio-technical environments<br>• Specify a shared manifesto (often based on the roles and value propositions of different stakeholders) |
| Ecosystem Maturity Map | Develop and assess an ecosystem development roadmap specifically for a national or local community or city. |
| Sector Development Canvas | • Develop and assess green and digital transformation in a specific sector.<br>• Monitor and evaluate cross-sectoral exchanges and flows of both supply chains and talent regeneration. |
| Storytelling Tool | • Create an engagement narrative for a project.<br>• Maintain sustainable relationship among stakeholders. |

Given the limited scope of this study, only the *Stakeholder Engagement Tool* is operationalized in the findings. The remaining tools constitute a future research resource.

## IV. RESEARCH FINDINGS

### A. Overview of the literature

The bibliometric data reveals a dynamic and maturing research landscape of 979 documents published between 1993 and 2026. An annual growth rate of 7.23% and an average of 13.6 citations per document show both sustained momentum and academic impact, underpinned by nearly 30,000 references reflecting three decades of scholarly conversation. The document average age of 10.7 years suggests a field mature enough to synthesize the Twin Transition dynamics.

TABLE III. DATASET MAIN INFORMATION

| *Category* | *Description* | *Results* |
|---|---|---|
| Documents | N | 979 |
| | Sources (Journals, Books, etc.) | 772 |
| | Timespan | 1993-2026 |
| Growth & Freshness | Annual Growth Rate (%) | 7.23% |
| | Document Average Age | 10.7 years |
| Impact | Average Citations Per Doc | 13.6 |
| | Total References | 29,529 |
| Document Contents | Author's Keywords (ID) | 2,630 |
| | Keyword Plus (DE) | 930 |
| Collaboration | Authors | 2,106 |
| | Co-Authors Per Doc | 2.65 |
| | International Co-authorships (%) | 20.63% |
| Document Types | Articles | 631 |
| | Conference Proceedings Paper | 348 |

The domain is highly collaborative and multidisciplinary. A co-authorship average of 2.65 per document and an international co-authorships rate above 20% reinforce GBS as a globally interconnected phenomenon. The prevalence of 2,630 unique author keywords — nearly triple the database-generated "Keywords Plus" count — indicates that researchers are moving beyond broad thematic descriptions toward highly specific methodologies, a sign of increasing intellectual specialization. The balanced mix of 631 journal articles and 348 conference papers underscores a vibrant living laboratory culture in which emerging ideas are rapidly shared at conferences before being codified in journals, capturing real-time industry shifts, such as the AI-driven infrastructure upgrades in "Middle Power" hubs. Together, these statistics validate GBS as a robust, globally-integrated area of study, though the high keyword specificity signals that academic discourse must now synchronize with the pace of practitioner-defined transformation.

### B. Conceptual structure and evolutions

To map the GBS conceptual landscape within the MLP-STT framework, this study applies a tripartite analytical approach: co-occurrence, factorial, and thematic evolution analysis. Co-occurrence analysis identifies the strength of direct thematic associations—revealing the "regime" of established knowledge and industry practice—while factorial analysis (Multiple Correspondence Analysis, MCA) uncovers the latent dimensions that distinguish mature, centralized themes from emerging "niche" innovations. Thematic evolution analysis then tracks concept trajectories of these concepts over time, validating whether isolated niches are gaining momentum or stagnating. Together, these findings answer how GBS centers are transitioning from labor-arbitrage models toward high-value orchestration hubs.

#### 1) Keyword co-occurrence network

The keyword co-occurrence network (Fig. 1) reveals five clusters: the Red Cluster anchors the traditional BPO and offshoring narrative, centered on risk and performance in established hubs like India and the Philippines. The Green Cluster represents the technical "engine room" of process engineering, utilizing simulation and process mining to drive optimization. The Blue Cluster signifies structural maturation, moving the domain from basic Shared Service Centers (SSCs) and accounting toward integrated GBS frameworks governed by transaction cost economics. The Yellow and Purple Clusters represent the macro-strategic shift toward the digital economy and the infusion of AI into supply chain capabilities.

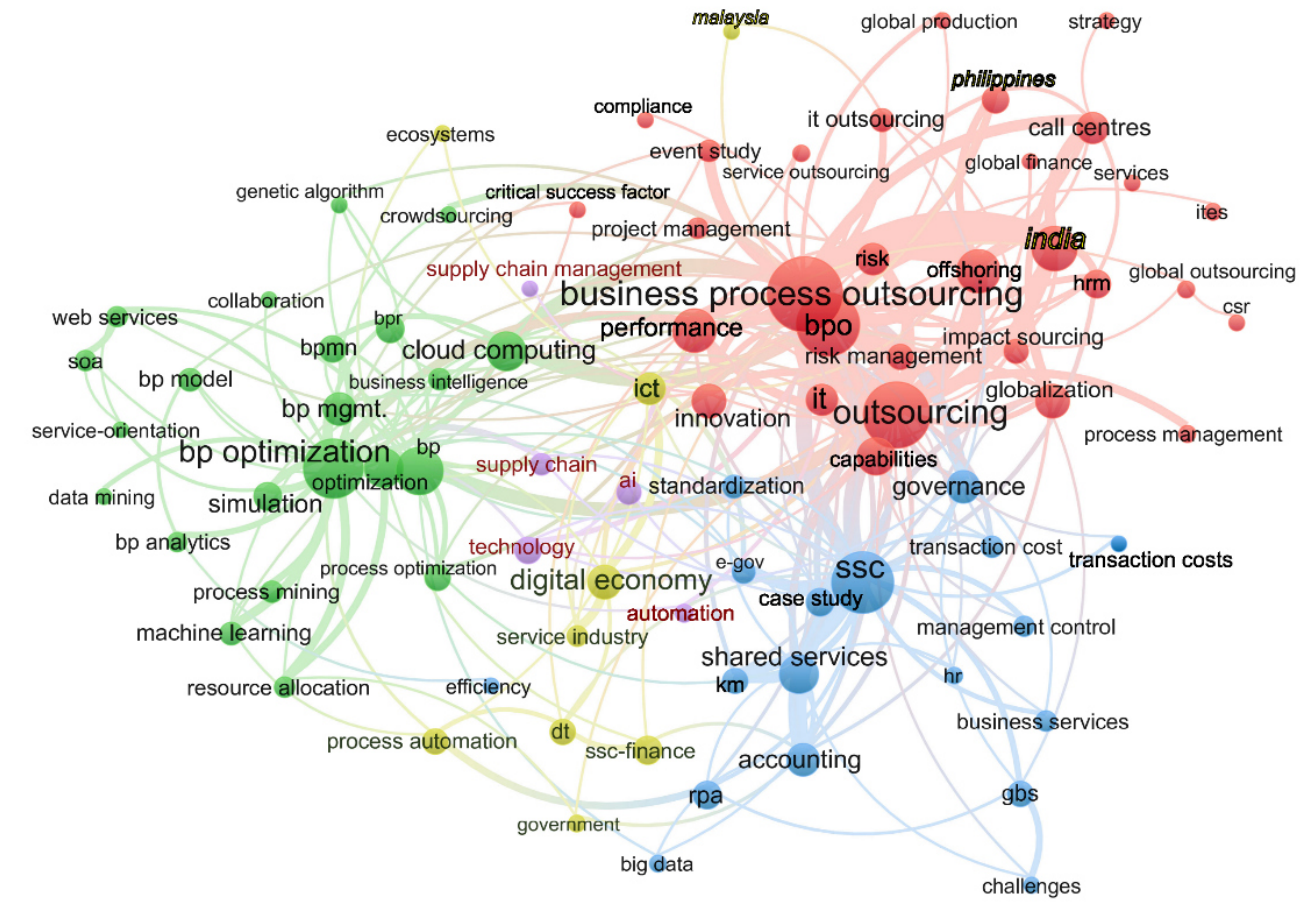

Fig. 1. Conceptual network: keyword co-occurrence visualization

Fig. 1 also contextualizes twin transition in the domain. The network's spatial structure is telling: the Green and Yellow Clusters occupy a bridging position between the legacy Red Cluster and the emergent digital economy nodes, visually encoding the transition from process efficiency to ecosystem intelligence. Terms such as *ICT*, *IT, innovation*, *digital economy*, and *process automation* appear at the gravitational center of the network, functioning as conceptual connectors between the BPO "regime" and the emerging AI-driven paradigm. Notably, *RPA* and *machine learning* appear in proximity to both optimization-oriented and digital economy clusters, confirming their dual role as operational tools and strategic enablers of the green digital Twin Transition. The relative peripheral positioning of *GBS, Corporate Social Responsibility (CSR)*, *Malaysia*, and *government* — visible but not yet densely connected — signals relatively less attention over the themes. It also reveals the importance to reposition GBS as a central pillar of national and corporate digital transformation strategies.

### 2) *Factorial Analysis*

The factorial map (Fig. 2) is dominated by a large central red cluster, where terms such as *process automation, cloud computing, and performance management* converge. This red cluster represents the stabilized core of the GBS domain, where technical efficiency and standardized data management have become the industry baseline. The proximity of RPA and process automation highlights the focus on automated processes, while the co-location of cloud computing and digital economy reflects their interplay as technological and economic drivers. Peripheral to this core, smaller isolated clusters carry distinct implications. IT-enabled services (ITES) form a specialized green cluster in the lower-left, less integrated with the broader field. *Public administration* and *process innovation* occupy a purple region in the upper area, indicating a separate policy-oriented discourse. *E-government* bridges public administration and the central technological cluster in the upper-middle section. The statistical isolation of these public and policy clusters reveals a structural gap: while the technical subsystem of GBS is highly integrated, its social and governance subsystems —essential for national innovation ecosystems— remain conceptually decoupled in the academic literature.

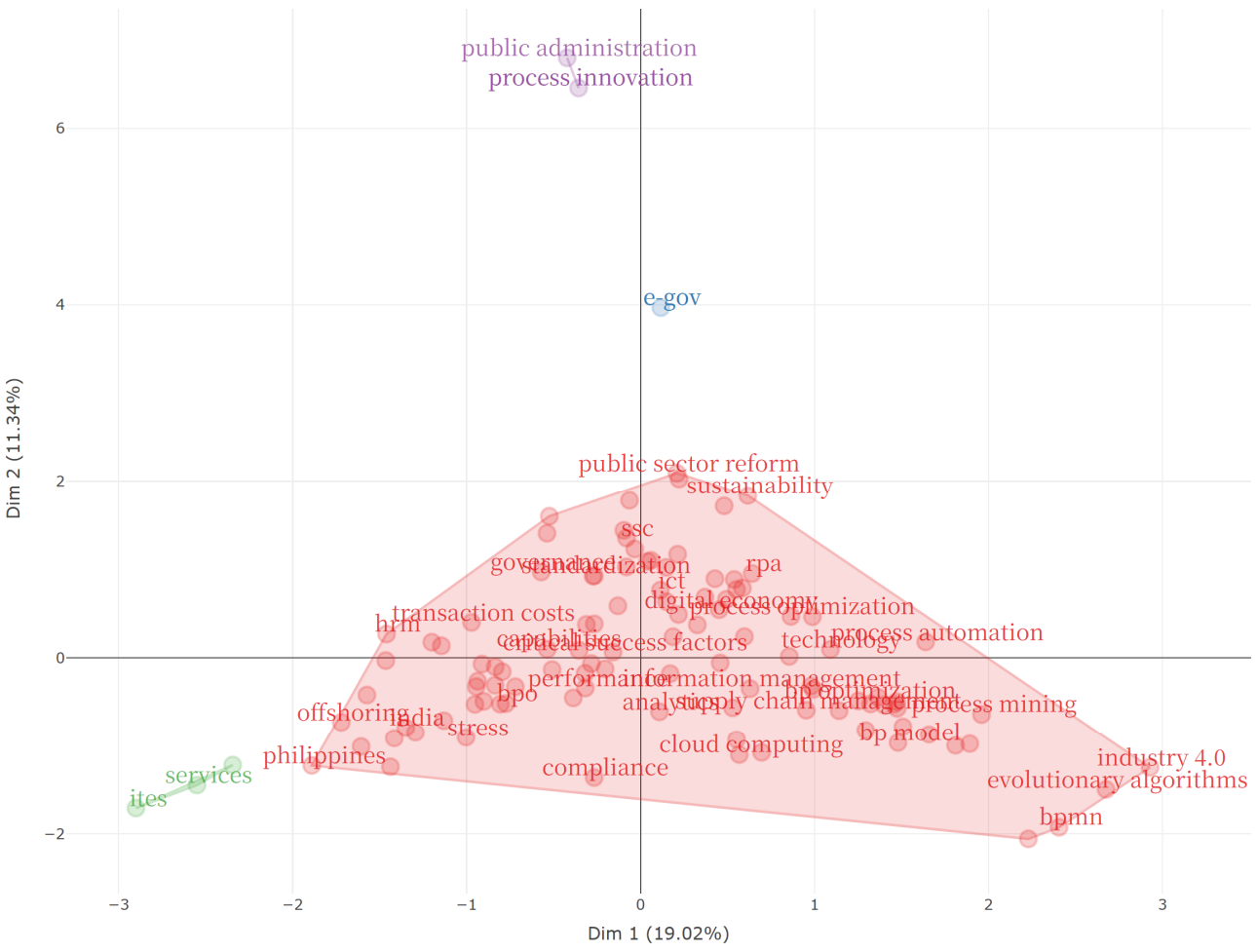

Fig. 2. Factorial Analysis

A critical reflection suggests the existence of "conceptual inertia." The density of the central cluster may mask a "digitalization paradox" in which traditional efficiency metrics (RPA, cost-reduction) overshadow the emerging Twin Transition requirements of green ICT and human-centric AI. Furthermore, the isolation of niche clusters like *IT-enabled services* (ITES) and *process innovation* indicates that the GBS 5.0 paradigm—while active in policy reports like Malaysia's GBS 5.0 initiative—has yet to reach critical mass in peer-reviewed synthesis, signaling a significant opportunity for future research.

### 3) *Thematic evolution analysis*

Grounding the roadmapping, a longitudinal analysis identifies the shift of clusters from 'niche' to 'motor' regimes using the MLP-STT framework. Fig. 5 illustrates thematic evolution through a longitudinal lens across three periods and four subfigures. Fig. 5(a) shows that BPO, Business Process Optimization, and SSC dominated the early period (1993–2015) as the established GBS "regime," focused on efficiency, cost reduction, and task centralization. The middle period (2016–2020) introduced RPA, performance metrics, and ICT as bridging themes. The current period (2021–2026) shows significant diversification, enriched by Digital Transformation (DT), CSR, process automation, and the digital economy.

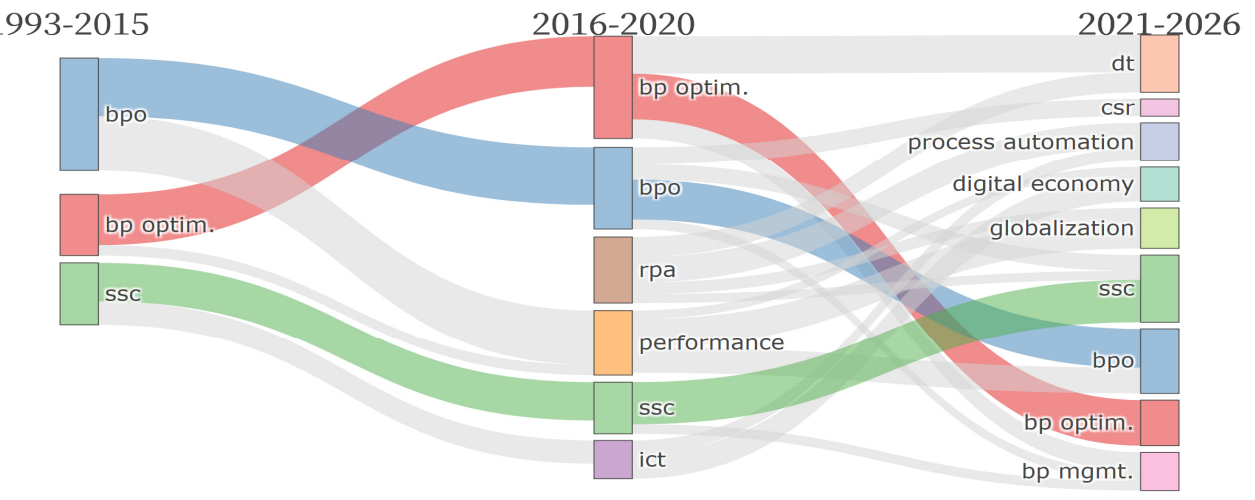

Fig. 3 Thematic evolution (Sankey) map: Showing the longitudinal flow and "evolution" or "mutation" of keywords across three periods.

Figs. 5(a)–(c) map each period onto four quadrants defined by centrality (x-axis) and density (y-axis) [45]:

- *Motor Themes* (upper-right): highly developed and essential to the field.
- *Niche Themes* (upper-left): internally well-developed (high density), but with limited cross-cluster interaction (low centrality).
- *Emerging or Declining Themes* (lower-left): weakly developed and marginal, usually either just entering the literature or losing significance.
- *Basic Themes* (lower-right): transversal and foundational topics, but lacking specialization.

Through the STT-MLP lens, *Motor* and *Basic* themes function as the "engines" and "regimes" that stabilize the GBS landscape. Fig. 5(a) shows India and outsourcing anchoring the BPO regime in labor arbitrage realities. Fig. 5(b) confirms Cloud Computing as a key *Motor* theme alongside accounting and shared services, marking a socio-technical upgrade from physical centralization to digitalized infrastructure, with "offshoring" joined "outsourcing" as a *Basic* theme. Fig. 5(c) shows "capabilities" as a key *Motor* theme, while DT, AI, and Machine Learning consolidate as *Basic* foundational themes, signaling a trend toward algorithmic intelligence. *Niche* and *Marginal* themes offer a basis for predicting the next five-year cycle. Transaction Cost was an early *Niche* theme, reflecting theoretical isolation; Governance appeared as a marginal theme in the middle period (Fig. 5(b)), as global networks grew more complex to manage. In the current period, Blockchain, ESG, and Employee Health occupy the *Niche* quadrant, as specialized tracks yet to achieve *Motor* status.

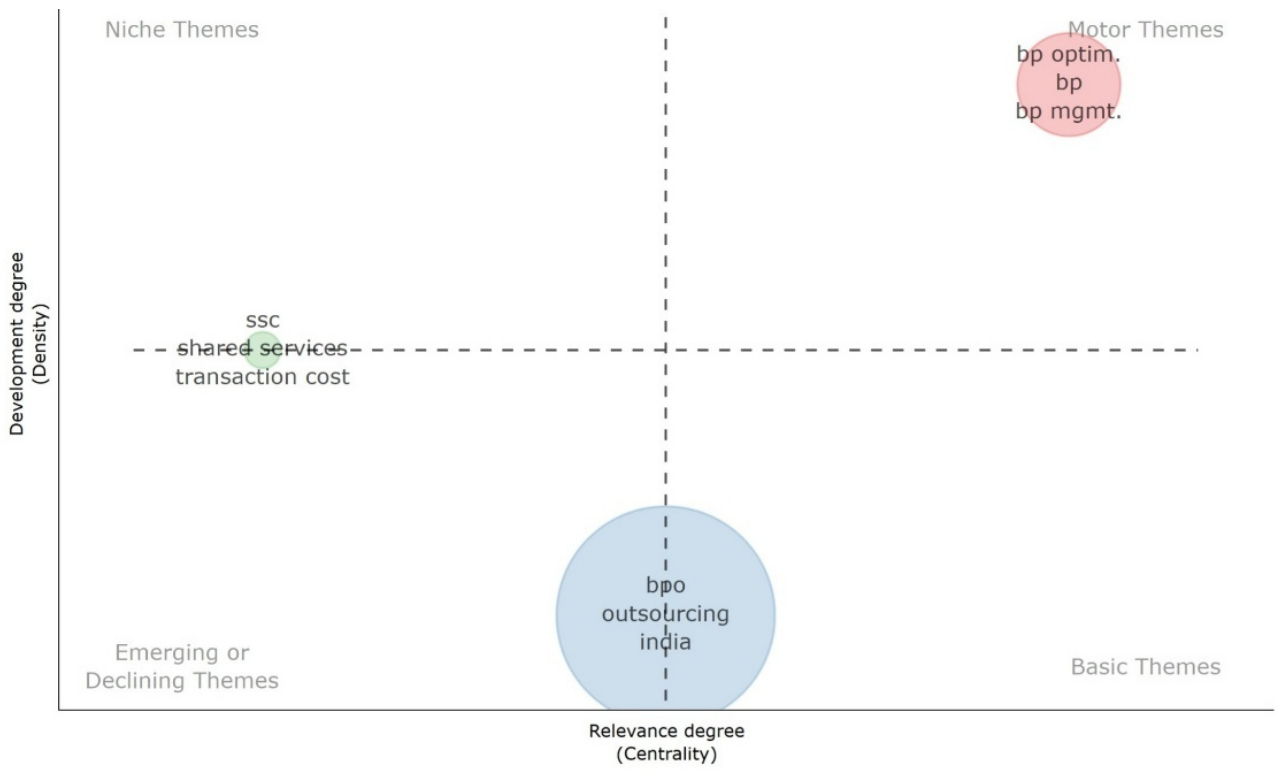

(a) **Strategic map (1993–2015):** The foundational era dominated by BPO and early outsourcing, especially related to India and transaction cost.

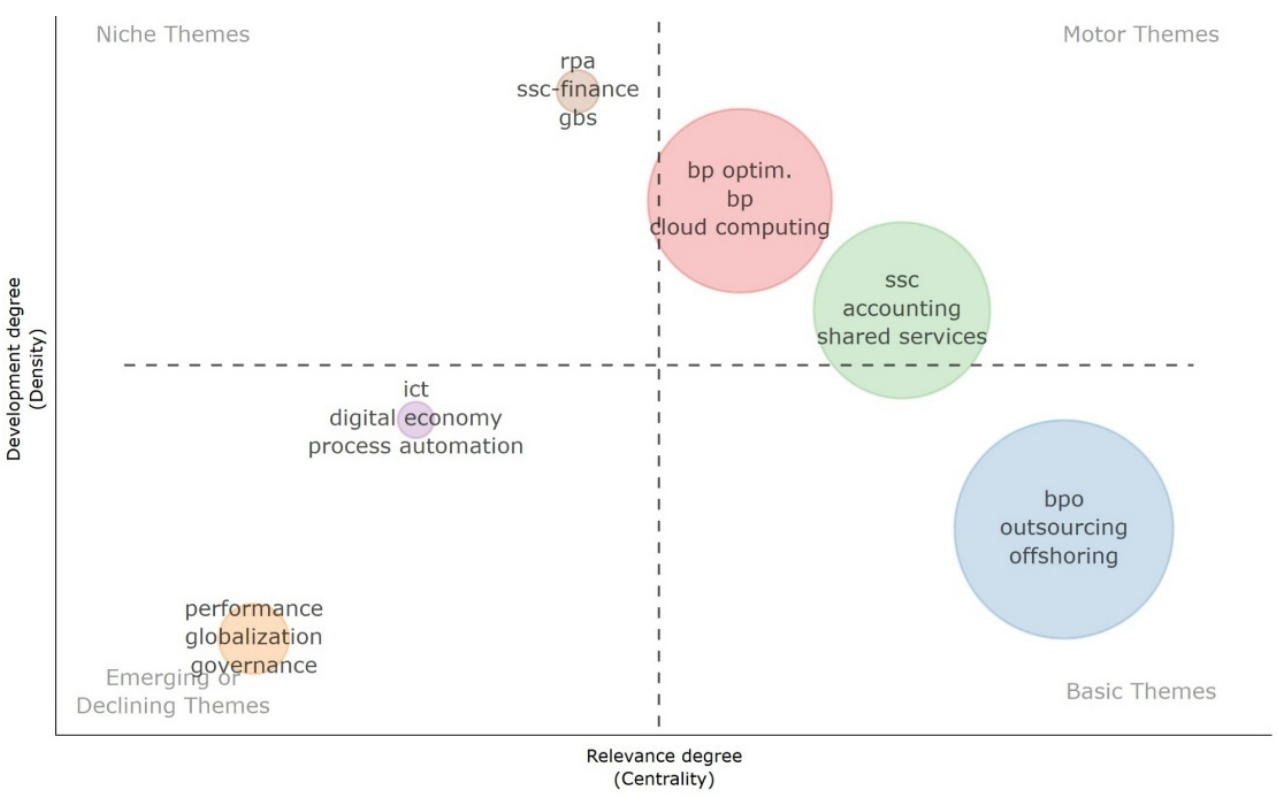

(b) **Strategic map (2016–2020):** The transitional era featuring the rise of RPA, ICT, cloud computing and process automation.

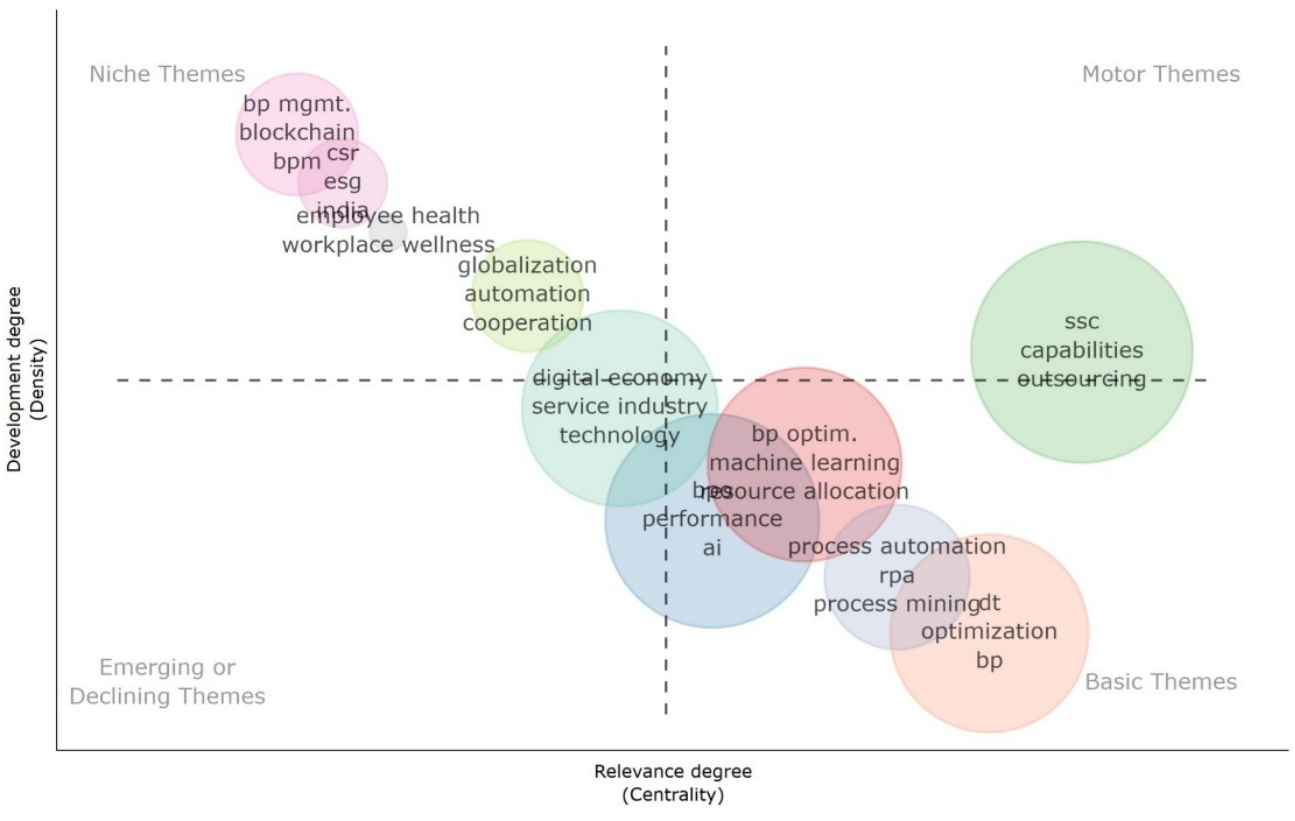

(c) **Strategic map (2021–2026):** The current era characterized by AI, digital transformation and CSR.

Fig. 5. Thematic evolution and strategic mapping of the GBS research landscape (1993–2026).

Critically, the positioning of Process Mining and AI near the boundary of the *Basic* quadrant marks them as "emerging engines." It is projected that in the next five years, the current niche of ESG will merge with Machine Learning to form a new "Sustainable Intelligent GBS" regime. Altogether, the "regime" has become more sophisticated in terms of domain expertise and analytics: evolving from transaction costs, toward accounting, performance, and governance that are contextualized in the socio-technical transitions of process automation and mining, ESG requirements, AI, resource allocation and service delivery planning. Overall, the field has shifted decisively from cost-reduction toward value-driven organizational capabilities.

## C. *Most relevant countries*

Fig. 4 reveals a stratified geographic distribution of research output. The collaborative network points toward the growing role of "Middle Power" innovation hubs alongside the dominant outputs of the United States and China.

Fig. 4 (a) identifies the United States and China as the primary centers of document production. A secondary tier — including India, the United Kingdom, Germany, the Netherlands, and Australia— functions as the "connective tissues" of the network, maintaining high output across diverse thematic areas. Fig. 4 (b) maps the "social structure" of the domain: the United States holds the most extensive global links, while China shows robust connectivity with Asian neighbors. The 20.63% international co-authorship rate underscores a field that rejects isolationism. Poland, France, and Malaysia emerge as regional knowledge production sites, with similar patterns evident among West Pacific Rim countries including the Philippines, Japan, and South Korea.

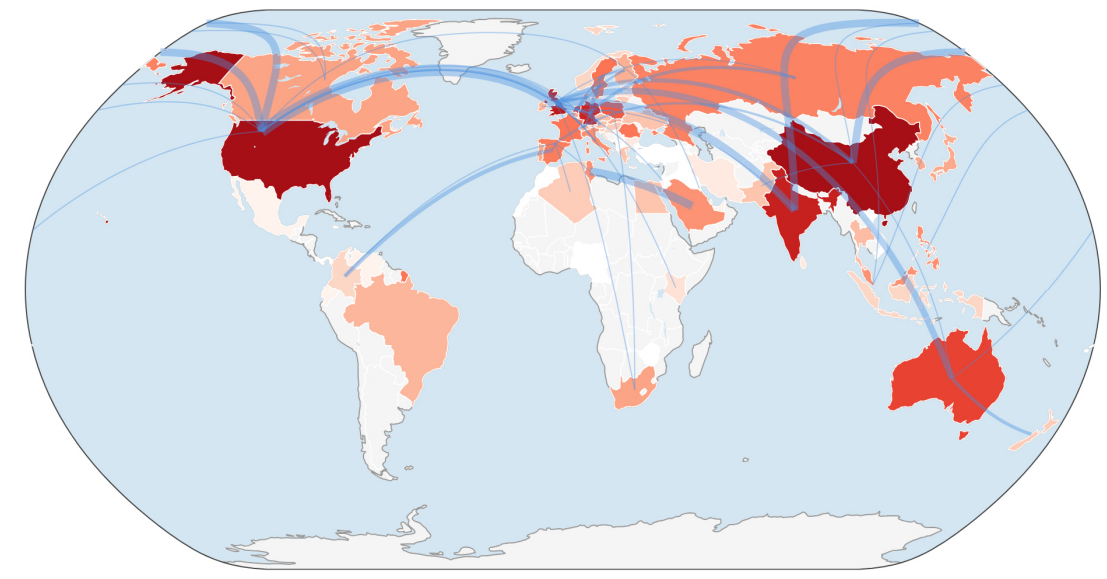

(a) World Map of International Collaboration

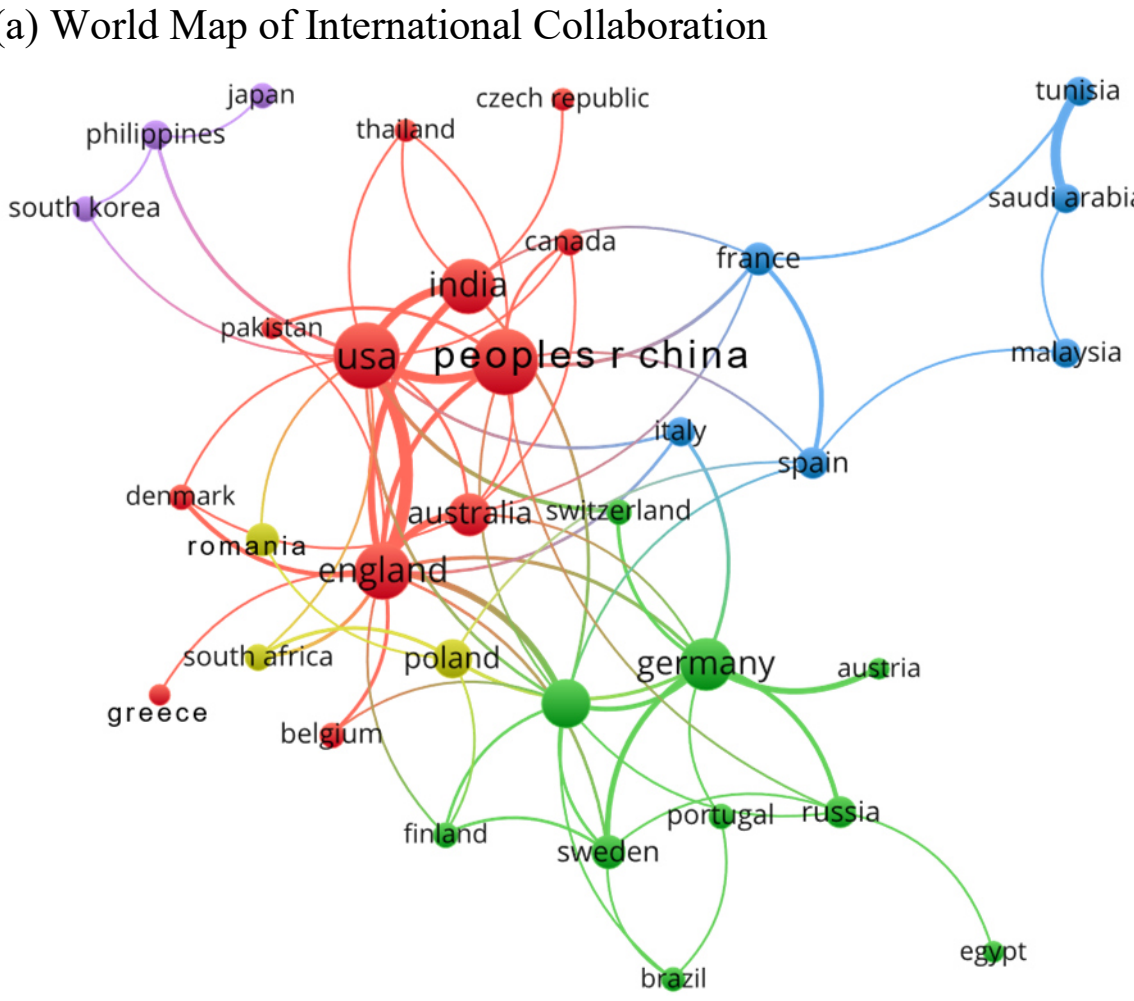

(b) Network Visualization of Co-authorship

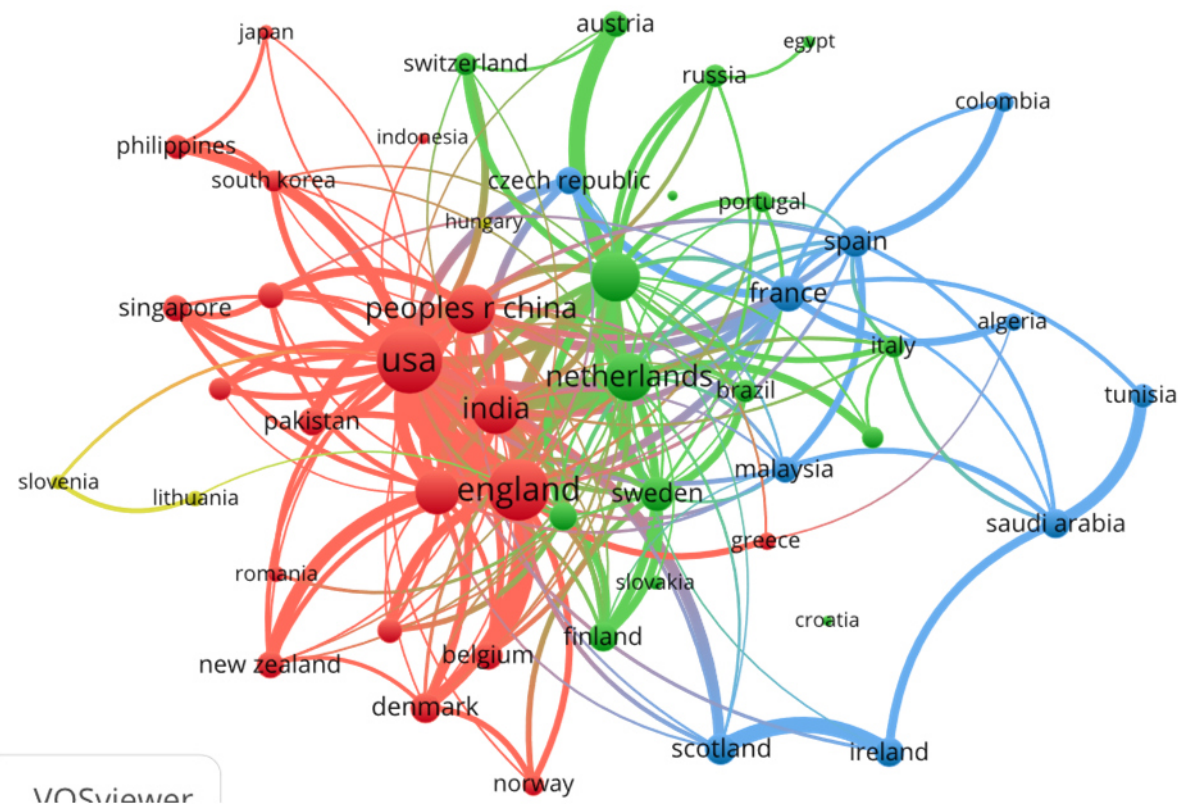

(c) Bibliographic Coupling Visualization

Fig. 4. Top relevant countries

Fig. 4 (c) measures intellectual proximity through bibliographic coupling, confirming higher degrees of reference sharing than co-authorship alone. Poland and Portugal emerge as critical bridge nodes within the European cluster, while Malaysia serves as a vital gateway for the ASEAN-Pacific research nexus. When compared to Fig. 4 (b), this network topology suggests a paradigm shift from a "hub-and-spoke" model of labor to a "distributed intelligence" model of collaboration —a trend essential for GBS leaders orchestrating resilient, multi-regional supply chains.

## V. DISCUSSION

The bibliometric evidence confirms a profound transition within the GBS domain. Thematic evolution aligns with broader innovation ecosystems, reflecting a responsible societal shift toward skill and service upgrades, but not yet the socio-technical transition toward full sustainability.

### A. *Answering Research Question*

The field is moving away from the traditional "regime" of discrete Business Process Outsourcing (BPO) and Shared Service Centers (SSC) toward highly integrated, digital ICT-centric ecosystems. Efficiency and optimization remain foundational, but are now subsumed by the strategic imperatives of Twin Transition, innovation ecosystems, and the digital economy, — elevating the role of *ecosystem orchestration*. The rise of RPA and ICT as bridging themes represents a functional shift: AI-native workflows transform routine-based delivery to technologically-augmented models, reaffirming GBS as a strategic value driver that reshapes how global work is organized and valued.

Fig. 1 and Fig. 2 highlight that the central nodes of ICT, Innovation, and AI have become the primary bridges linking traditional BPO and SSC clusters, the key terms for modern policy, regulation, and standards. These GBS units now act as an "operational airlock" where landscape-level pressures—such as the EU's dual mandate and Carbon Border Adjustment Mechanisms—are translated into standardized, sustainable workflows. [9]. This validates the study's central hypothesis: GBS has evolved from a cost-arbitrage instrument into an innovative ecosystem orchestration mechanism, operating at the intersection of technological, regulatory, and organizational transformation: By utilizing bibliometric science mapping as an empirical foundation, this transition offers a structured, evidence-based service design that provides high-quality service environments for multinational corporations (MNCs) and their global partners.

### B. *Match and Contribution*

GBS nodes serve as ideal living laboratories for Socio-Technical Systems (STS) theory and the MLP-STT framework, synthesizing the *technical* (process optimization and AI), the *social-organizational* (governance and CSR), and the *spatial* (SSCs as physical and digital entities) into a resilient operational ecosystem. Originally envisioned for public policy stakeholders, the ITU toolkit demonstrates its novel utility here for MNCs navigating landscape pressures (such as carbon neutrality and AI disruption), effectively evolving GBS practices into scalable niche innovations.

Fig. 5 confirms that ESG themes have become the operational core of Digital Transformation. Emerging themes — business analytics, Service-Oriented Architecture (SOA), and process mining — act as "functional connective tissue" linking operational processes to higher-order concepts such as autonomous workflows and innovation ecosystems, pointing toward a *Sustainable Intelligence* model. This shift addresses the need for both human-centric (talent) and logistics-centric (supply chain) realism within an Industry 5.0 context, balancing high-tech optimization with socially responsible practices. Workplace wellness and employee health now emerge as "motor themes," signaling that human capital is a core driver—not just a support function—of GBS competitiveness.

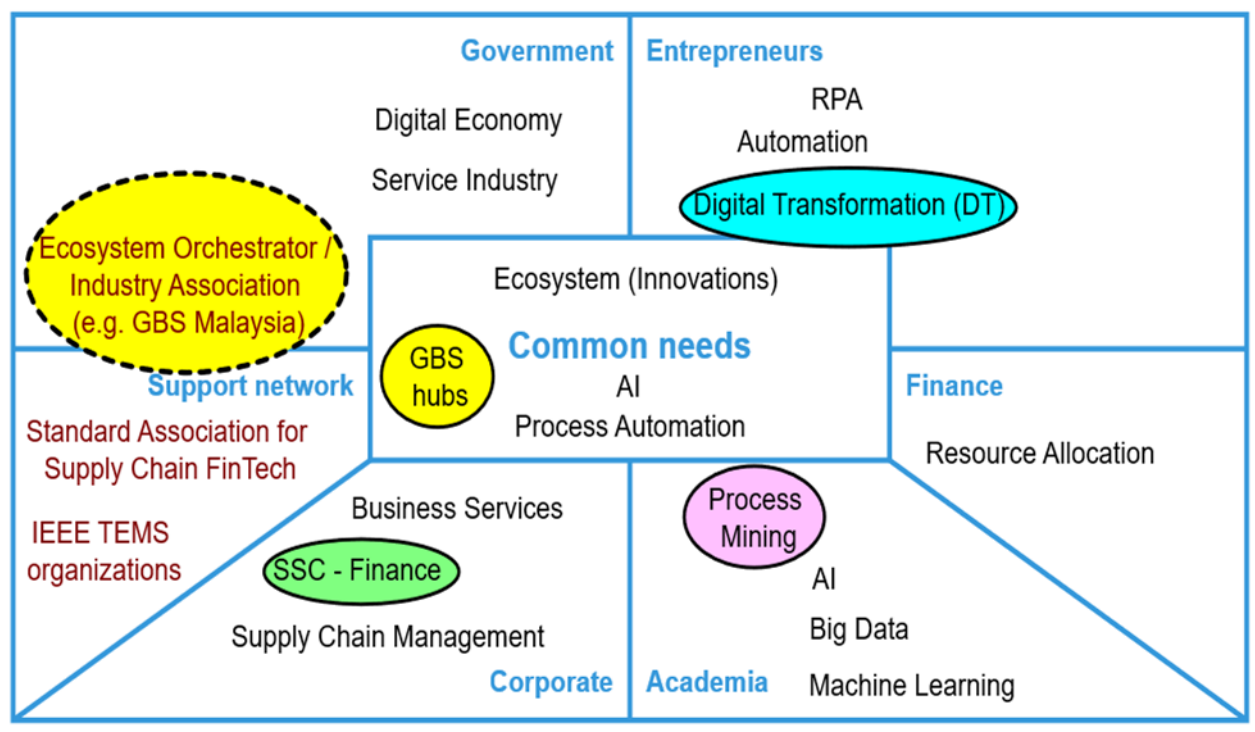

Fig. 6. Stakeholder Engagement Tool Canvas

Integrating bibliometric science mapping with the MLP-STT framework provides a robust stakeholder engagement foundation, showing how industrial policies act as regime-level influences compelling MNCs to adapt their roadmaps. The Stakeholder Map (Fig. 6) further validates this contribution by revealing four threads, leveraging *AI and process automation* for innovation to solve common needs.

*The Policy-Compliance Thread (Government to Corporate)* translates macro-level mandates (e.g., CBAM, CSDDD) into corporate compliance standards, elevating GBS from cost centers to guardian of the MNC's "license to operate" under the EU dual mandate.

*The Talent-Innovation Thread (Entrepreneur to Academia)* bridges entrepreneurial demand with academic research, enabling "Middle Power" hubs in Poland, Portugal, and Malaysia to capture higher-value delivery — with D&I strategies and hybrid work models serving as structural enablers for data-driven talent decisions.

*The Reward-Orchestration Thread (SSC-Finance & Support Networks*) evolves SSC-Finance into Supply Chain Finance, integrating green FinTech, RegTech, and automated auditing to financially incentivize sustainable behaviors across global value chains, ensuring data-driven logistics serve planet-centricity and enriching the Industry 5.0 landscape.

*The Living Lab Thread (Entrepreneur to Corporate)* democratizes innovation by providing entrepreneurs a structured environment to test AI-native workflows before rolling them out company-wide. Using MLP-STT and Industry 5.0 conceptual frameworks, this mechanism actively alleviates the technical and information asymmetries historically enjoyed by MNCs, recasting workforce reskilling as a strategic game played out across regime-niche dynamics.

## VI. CONCLUSION

This study confirms that Global Business Services (GBS) has evolved into a "living laboratory" for socio-technical systems of AI and human resources, offering an optimized developmental pathway within digital production networks.

## A. Limitations

Despite the high-value transformations in "Middle Power" hubs, a gap persists between the literature and empirical reality. Science mapping reveals a domain still heavily anchored in BPO models, creating a visibility gap for emerging leaders. Additionally, phenomena such as Global Capability Centers (GCC) were excluded from this scope. Future research must examine how these nations leverage GBS or GCC to bypass the middle-income trap and ensure their hosted GBS or GCC hubs remain functional despite a bifurcated cloud, likely by providing resilient, green-certified, and high-talent services in GVC. Another limitation is the absence of temporal burst detection (e.g., Kleinberg's algorithm) in the longitudinal analysis. Incorporating burst analysis would offer an early warning system enabling GBS leaders to identify weak yet **niche** signals — in generative AI ethics or carbon-border policies, for instance —before they stabilize into regulatory regimes.

## B. Concluding Remarks: GBS as a Structural Possibility

This study advances three interconnected contributions to technology management literature. First, it empirically validates the transition of GBS from a cost-arbitrage model to an ecosystem governance mechanism, confirmed via longitudinal bibliometric mapping. Second, it operationalizes the Sustainable Intelligence model, demonstrating that ESG themes have moved from the analytical periphery to the operational core of digital transformation within GVC networks. Third, it establishes the GBS hub—particularly in "Middle Power" locations—as a theoretically grounded living laboratory for the Twin Transition.

A cross-cutting insight emerges from this synthesis: the Twin Transition is not uniformly distributed. It concentrates in hubs where three capabilities converge — advanced digital infrastructure, regulatory alignment, and people-centered organizational culture. Crucially, while unguided digital labor platforms risk restricting localized upgrading [12], institutionalizing cross-border power asymmetries [13], and polarizing high-value remote work [14], orchestrated GBS hubs offer a structural possibility to counter these planetary network inequities. Within these hubs, HR functions acting as strategic business partners, data-driven talent pipelines, and trust-based hybrid models build the organizational substrate on which AI transformation securely runs.

Generalizing this model, three principles emerge for resilient GBS governance:

*1)* ***Co-dependence:*** Human capital strategy and AI adoption must function as simultaneous, co-dependent investments rather than sequential tasks

*2)* ***Holistic Design:*** The Twin Transition is a deliberate socio-technical realignment of GVC—not merely a technical upgrade—, in which human and algorithmic components must be designed together from the outset.

*3)* ***Disciplinary Reframing****:* GBS governance *i*s an inherently socio-technical discipline, *transcending its* operational or logistical *constraints*.

## C. Future Work and Perspectives

The integration of bibliometric data with the ITU ICT-Centric Innovation Ecosystem Toolkit provides a rigorous framework for mapping the Twin Transition as a socio-technical realignment projects across global division of labor. While this paper uses the ITU toolkit to map GBS nodes as operational airlocks, future work will expand this framework into a comprehensive SoS architecture across GVC.

Our vision for the Safe-and-Sustainable-by-Design (SSbD) Chips Value Chain Services (SCVCS) Research Initiative [46], we aim to investigate the extent to which GBS units might function as the core socio-technical control tower for advanced digital economies. Specifically, we propose a research trajectory that seeks to leverage global GBS talent and AI orchestration to explore the integration of:

*1) AI and semiconductor roadmaps* for regeneratively accountable and environmentally safe supplies;

*2) Cognitive Computing Continuum* for controlling loops that balance yields with resource circularity;

*3) Trustworthy autonomous AI agents*, serving as regulatory proxies to secure RegTech innovation using industrial data spaces; and

*4) Regenerative socio-technical architecture* to support material lifecycle tracking using digital product passports.

By deploying the remaining components of the ITU toolkit—the *Ecosystem Maturity Map*, *Sector Development Canvas*, and *Storytelling Tool*—this initiative will document the GBS transition from outsourcing destinations to innovation hubs. Redefining GBS as a primary socio-technical mechanism for navigating AI and global talent flows remains essential for maintaining multinational competitiveness within a decoupled global economy and supply chains.

## Acknowledgment

The authors gratefully acknowledge the constructive feedback of the double-blind peer reviewers whose scrutiny strengthened this work; all remaining errors and interpretations are solely the authors' own responsibility.